# Gold nanocone array with wide angles and high absorptivity for light absorber

*Xu Jiawei\* Huang Ran Tao Lei Zhang Jiaming Wu Xuanxuan Yang Pengchong Lyu Shuangbao\* Duan Jinglai*

*Xu Jiawei and Huang Ran contribute equally in this article*

Dr. J. W. Xu, Dr. L. Tao, J. M. Zhang, X. X. Wu, P. C. Yang, Prof. J. L. Duan
Advanced Energy Science and Technology Guangdong Laboratory
516000 Huizhou, China
Xujiawei@gdlhz.ac.cn
Dr. R. Huang, J. M. Zhang, P. C. Yang, Dr. S. B. Lyu, Prof. J. L. Duan
Institute of Modern Physics, Chinese Academic of Sciences
730000 Lanzhou, China
Lyushuangbao@impcas.ac.cn
J. M. Zhang, P. C. Yang, Dr. S. B. Lyu, Prof. J. L. Duan
School of Nuclear Science and Technology, University of Chinese Academic of Sciences
100049 Beijing, China
X. X. Wu
College of Physics Science and Technology, Hebei University
071002 Baoding, China

The gold nanocone array, fabricated with the ion track method, exhibits exceptionally high light absorptivity, making it a wide angle light absorber. The finite difference time domain (FDTD) method was employed to study the influence of the geometric parameters on total absorptivity and promote strategies to enhance the absorptivity of the gold nanocone array. Experimental measurements were concluded to assess the total absorption and bidirectional reflection distribution function (BRDF) properties of the gold nanocone array. The absorptivity measurement results demonstrate that the gold nanocone array exhibits a wide angle (5-75°) of incidence and high absorptivity properties, achieving an absorptivity over 0.98. The BRDF results show that the reflection of the gold nanocone array is predominantly attributed to non-specular reflections, and exhibits excellent Lambert properties. This study reveals the influence of structural parameters and light incidence angle on the absorptivity of the gold nanocone array, which plays an important role in further understanding the light absorption/reflection mechanism and enhancing the absorptivity.

## 1 Introduction

The nanocone array is a typical light trapping structure, providing a natural advantage to achieve broadband high absorptivity.[1–5] It creates an effective medium between the substrate and the air, with the refractive index gradually changing from the air to the substrate[6]. The nanocones have a non-complete symmetry, and its diameter varies continuously with its height, which causes it to resonate in multiple wavelengths, facilitating multi-peak resonance and broadband absorption through superposition. These characteristics position the nanocone array for potential applications in various fields, including solar cells[7–9], photocatalysis[10–12], infrared detection[13, 14] and so on.
The materials and geometry of the nanocone array are essential for studying light absorber [1, 15, 16]. Researchers have studied nanocone structures of metals, non-metals, oxides, semiconductors, and more. Metal nanostructure arrays, such as gold nanocone array[1, 17, 18], tungsten nanocone array[16], silver nanocone array[5], with antireflective structures, exhibit surface plasmon resonance enhancement and are suitable for use as high-absorption materials such as blackbody materials and surface-enhanced Raman spectroscopy[1, 18]. Mana Toma fabricated a flexible gold nanocone array film with excellent antireflective and high light-absorption properties.[1] The geometric parameters of the nanocone array has a significant impact on its absorptivity, and the maximum absorptivity and absorption band can be obtained in a silicon nanocone array, and its absorptivity can reach more than 0.99, which has a wide range of applications in the field of solar cells.[9, 15, 19, 20] François Atteia[15] and Da Yun[16]conducted experimental measurements and theoretical calculations to investigate the influence of geometric parameters on the absorption of the nanocone array, determined the geometric parameters with the best absorptivity. Many studies have concentrated on the relationship between absorption/reflectance and wavelength. However, these studies have focused on the absorptivity at a specific angle, and the understanding of the effects and mechanisms of different incident and reflected light angles on the absorptivity is unclear. The bidirectional reflection distribution function (BRDF) defines the directional reflection characteristics of a material surface.[21] BRDF is a function closely related to the angle of incidence and reflection, that characterizes the reflective properties of materials in all directions of space. To be precise, BRDF is related to the material roughness, light wavelength, incident angle and other factors.[22]Black silicon, a typical nanocone array, exhibits strong reflection in large angles,[15] and shows significant reflection in specular angles.The study of BRDF of the materials is expected to promote their applications in aerospace remote sensing[23], geological surveying, precision guidance, target detection, and target simulation.[24] BRDF research can reveal the reflective properties and mechanisms of light-absorber, and can also promote their application in various fields, but the understanding of BRDF of light absorber is still unclear. In this work, we improve the broadband absorptivity by strategically adjusting the geometric parameters of the gold nanocone array, utilizing Finite Difference Time Domain (FDTD)[25, 26] simulations. We successfully fabricated a gold nanocone array with wide angle high absorptivity using the ion track method. Additionally, we conducted experimental measurements to explore the reflective properties and underlying mechanisms of the gold nanocone array by hemispherical reflectance (or absorption) and BRDF.

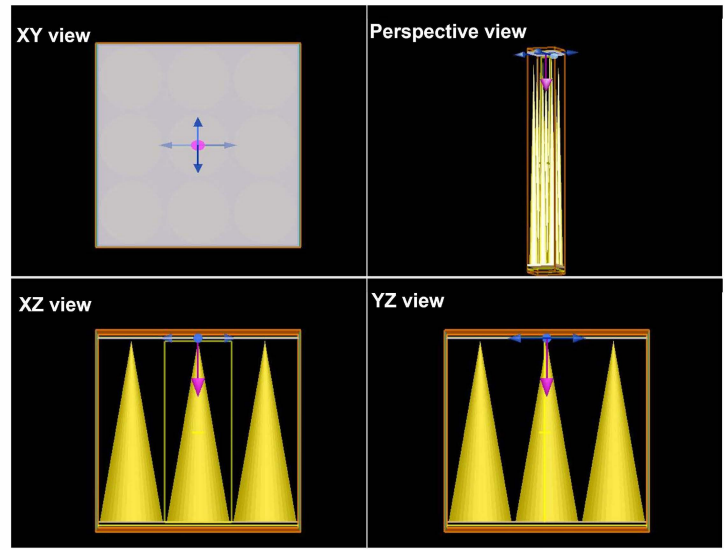

Figure 1: Gold nanocone array model constructed by FDTD

## 2 Results and Discussion

### 2.1 Influence of geometric parameters

In fact, the absorptivity of gold nanocone array depends on their geometric parameters,[1, 15, 16] such as cone angle, density, and height. Therefore, we use FDTD to investigate the correlation between these geometric parameters and the absorptivity, aiming to find the maximum absorptivity. In the simulation, each geometric parameter was systematically modified in isolation, allowing for a detailed evaluation of the impact of each parameter on the absorptivity. Figure 1 shows the gold nanocone array model simulated in this paper, comprising a gold nanocone array on a gold layer substrate.

The first geometric parameter is the cone angle with the variety from 1.4° to 2.6°. The calculated results are shown in Figure 2(a) with distinct colored solid lines. In this Figure, a plateau can be observed be-tween 0.3 and 0.6 $\mu m$ at all angles, where the absorptivity is approximately equal to 1. Following this plateau, there is a gradual decrease in absorption as the wavelength increases. Notably, the high absorptivity plateau remains consistent across all cone angles calculated.
It is obvious that the absorptivity at longer wavelengths increases with the increment of the cone angles. Specifically, the absorptivity increases gradually as the cone angle rises from 1.4° to 2.3°, but there is a decrease at a cone angle of 2.6°. The evidence suggests that the large cone angle results in a higher absorptivity for long wavelengths. However, when the cone angle exceeds 2.3°, the adjacent nanocone's bottom begin to overlap, leading to a reduction in the effective height of the nanocones. Consequently, this reduction results in a decreased absorption rate of the gold nanocone arrays at longer wavelengths.
The second geometric parameter is the height of the cone, which ranges from 10 to 25 $\mu m$. Figure 2(b) presents the calculated absorption spectrum of gold nanocone array at various heights, with a cone angle of 2.3° and a density of 1e8 cones/cm$^2$. As shown in Figure 2(b), the absorptivity increases with the increasing of the height. When the nanocone height is higher than 15$\mu m$, there exists a plateau between 0.3 and 0.6 $\mu m$ followed by a gradual decrease in the longer wavelengths, and the plateau broadens with the height.
The third geometric parameter is the density of the nanocone array, which varies in the range of 1e7 to 1e8 cones/cm$^2$. Figure 2(c) shows the absorptivity of the gold nanocone array with a cone angle of 2.3° and a height of 25 $\mu m$ calculated in different densities. As shown in Figure 2(c), the gold nanocone array with a higher density exhibits higher absorptivity. Moreover, two peaks in absorptivity at wavelengths around 0.5 $\mu m$ and 0.7 $\mu m$ are observed in case that the density is below 5e7 cones/cm$^2$.
FDTD simulations indicate that in a gold nanocone array, higher cone height, larger cone angle, and higher density of nanocones lead to a higher absorptivity, under the condition that the nanocone bottom does not overlap. For the cone angle, a larger cone angle of the nanocone increases the area of the cone bottom that reduce the reflection from the base metal plane. However, the overlapping of adjacent nanocones reduces the actual cone height and increases reflection from the overlapping areas. With respect to the height of the nanocone, a higher height results in a larger bottom size, which helps enhance absorption across a wider wavelength range. The large bottom size of the nanocone also minimizes reflection from the base metal plane. Moreover, higher nanocone can produce more multiple reflections and absorption from their vicinity cone, further improving light absorption efficiency. Additionally, as the height of the nanocone increases, the change of the average refractive index between air and gold becomes more gradual, thereby mitigating Fresnel reflection[6]. We estimate the effective refractive index $n_{eff}$ for quadrilateral arrangements of cones using the effective medium approximation (EMA),[6] The effective refractive index of the gold nanocone array at heights of 10, 15, 20, and 25$\mu$m was calculated using the method mentioned above and shown in Figure 2(d). Figure 2(d) shows higher height of the nanocone has slower change of the $n_{eff}$. Regarding to the density of the nanocone array, as the density increases, the reduction of reflection from the base metal surfaces and simultaneously enhances multiple reflections among the cones, thereby improving the overall absorption efficiency of the gold nanocone array.

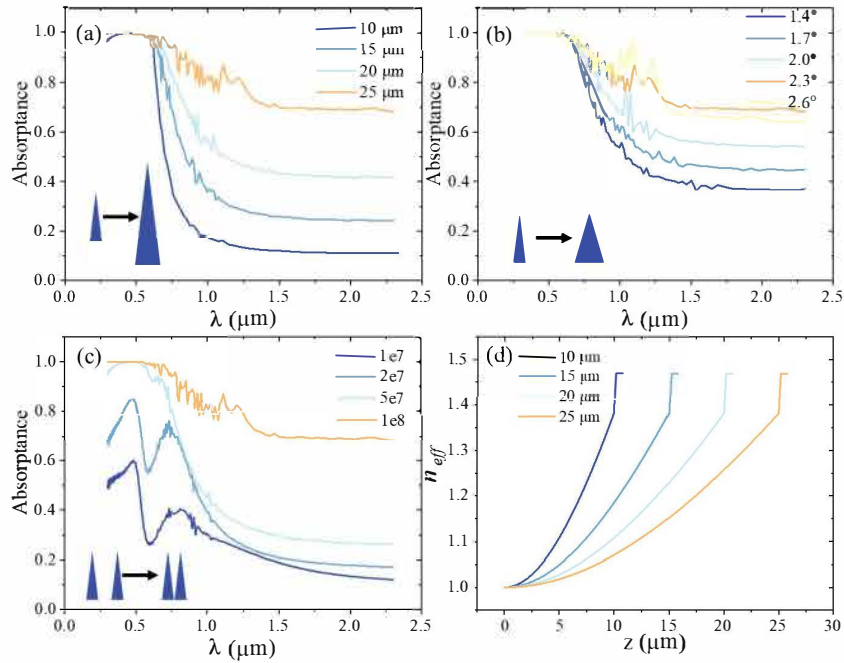

Figure 2: Absorption of FDTD calculations versus wavelength for gold nanocone array with different cone angles(a), height(b), density(c), and the $n_{eff}$ of gold nanocone array with different heights at 400 nm(d).

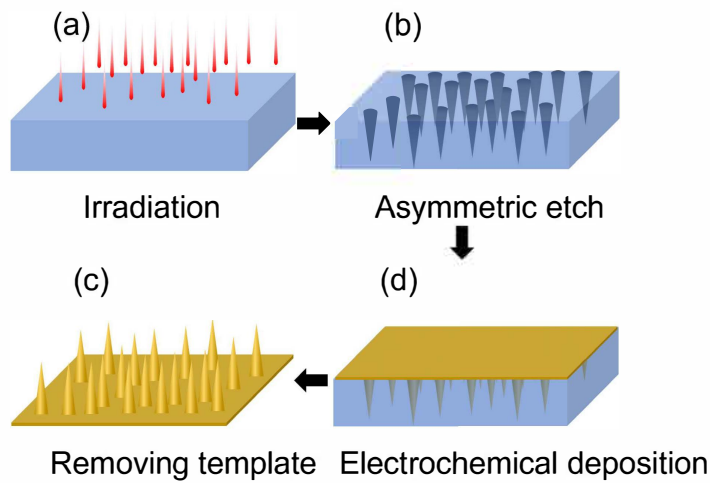

Figure 3: Schematic illustrations of the fabrication procedures of gold nanocone array.

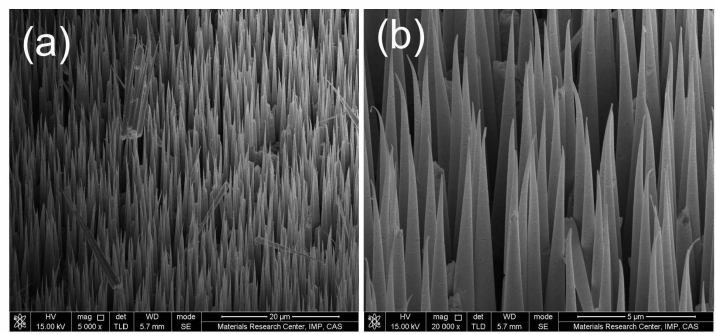

Figure 4: SEM images of the gold nanocone array.

## 2.2 Absorption of the fabricated gold nanocone array

The gold nanocone array were fabricated based on ion track method as shown in Figure 3. In brief, the main processes involved swift heavy ion irradiation of a polycarbonate (PC) foil, asymmetric etching, electrochemical deposition, and removal of the PC template. The structure of the prepared gold nanocone array, as shown in Figure 4, exhibited a uniform morphology and size, a standard conical shape, and a smooth surface with a clear contour.
The experimental results are presented as a thick solid line in Figure 5. Gold nanocone arrays exhibit excellent absorption rates of up to 0.98 and preserve this absorptivity in the wavelength range of 0.3-0.5 $\mu$m. The absorptivity gradually decreases after a wavelength grater than 0.5 $\mu$m and stabilizes after 1.5 $\mu$m, consistently exceeding 0.6 across the entire measurement range. Obviously, the gold nanocone array exhibits the distinct characteristic of broadband absorption.
The spectral absorbance of the gold nanocone array was simulated using FDTD and compared with the experimental measurement results. The simulated gold nanocone array with a height of 22 $\mu$m. As can be seen in Figure 5, the simulation results are in agreement with the experimental results, indicating the precision of the FDTD model and the calculation method for the gold nanocone array established in this paper. It is noteworthy to mention that actual gold nanocone areas are randomly distributed and the modeled array is set periodic. The random position of the nanocone in the array and the part broken nanocones in the sample are possible reasons for the slight discrepancy between the experimental measurements and the simulation results.

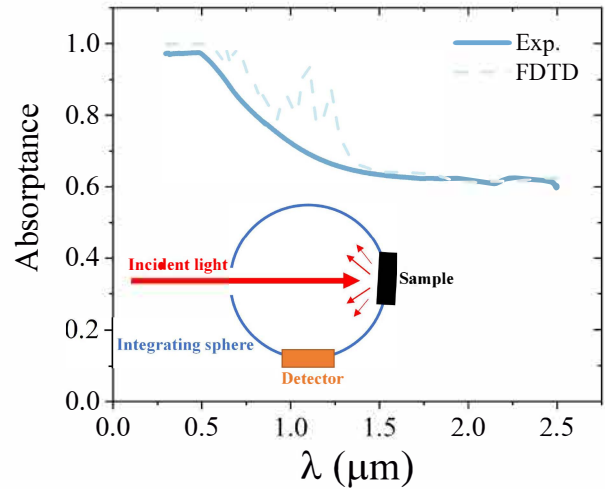

Figure 5: Experimental measurement and FDTD calculations of the absorption of the gold nanocone array.

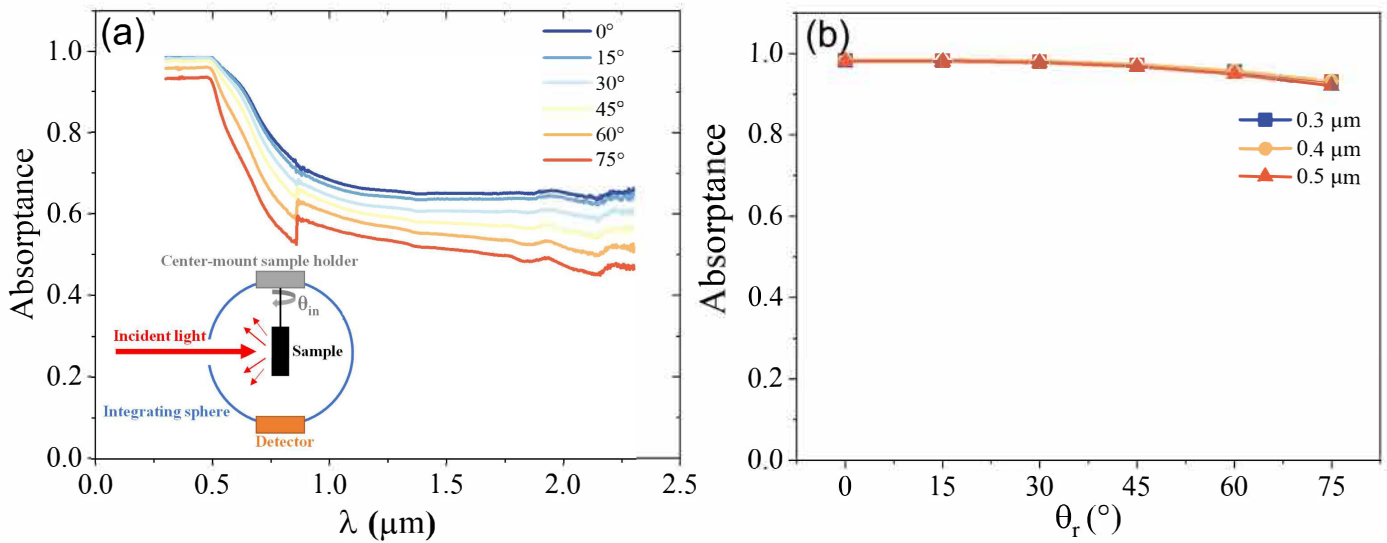

Figure 6: Influence of the incidence angle on the absorptivity of the gold nanocone array.

Additionally, we explore the dependence of absorptivity on the incident angles (5-75°), which is important for the development of practical wide angle absorber devices. As shown in Figure 6(a), the measured absorptivity of the gold nanocone array decreases as the incident angle increases, exhibiting similar characteristics to those of other nanomaterial absorbers[1, 15]. At an incident light angle of 75°, the highest absorptivity can reach 0.93 in the range of 0.3-0.5 $\mu m$, and still has an absorptivity rate higher than 0.5 in all measurement wavelength bands. Figure 6(b) shows a low dependence of absorptivity at the light incidence angle of the gold nanocone array, indicating that the gold nanocone array has good characteristics of high absorptivity at wide angles.

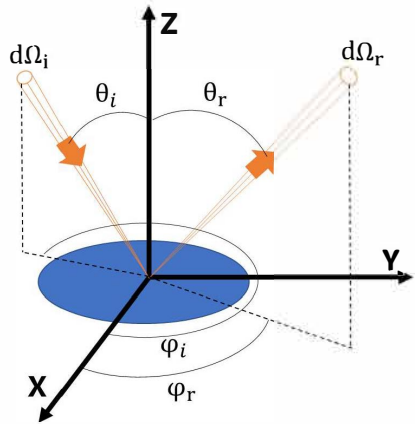

Figure 7: BRDF scattering geometry.

## 2.3 BRDF of gold nanocone array

We have demonstrated the properties of gold nanocone array on light capture. Additionally, we are also interested in the change in the absorptivity of light versus the angle of incidence and reflection.
In Figure 7, the geometric relationship of the BRDF measurement is illustrated, which is measured by measuring the ratio of the intensity of the reflected light to the incident light at any angle of the sample surface, as shown in Equations 1 and 2. The BRDF expressed by $F_T(\theta_i, \varphi_i, \theta_r, \varphi_r, \lambda)$, where $dL_r(\theta_i, \varphi_i, \theta_r, \varphi_r, \lambda)$ is the luminance of the sample in the direction $(\theta_r, \varphi_r)$, $dE_i(\theta_i, \varphi_i, \lambda)$ is the illumination of the sample surface in the direction $(\theta_i, \varphi_i)$. The $i, r$ represent the incidence and reflection, $\lambda$ is the wavelength of the measured light, $\theta$ and $\varphi$ represent the zenith and azimuth angles, respectively. It is important to note that the illumination source is an s-polarized laser.

$$F_T(\theta_i, \varphi_i, \theta_r, \varphi_r, \lambda) = \frac{dL_r(\theta_i, \varphi_i, \theta_r, \varphi_r, \lambda)}{dE_i(\theta_i, \varphi_i, \lambda)} \quad (1)$$

$$F_T(\theta_i, \varphi_i, \theta_r, \varphi_r, \lambda) = \frac{V_T}{V_S} F_S(\theta_i, \varphi_i, \theta_r, \varphi_r, \lambda) \quad (2)$$

$F_T(\theta_i, \varphi_i, \theta_r, \varphi_r, \lambda)$ and $F_S(\theta_i, \varphi_i, \theta_r, \varphi_r, \lambda)$ are the BRDFs of the target sample and the standard sample, respectively, and $V_T$ and $V_S$ are the measured signal values of the detectors of the target sample and the standard sample, respectively.
In order to investigate the impact of incident angle on the absorptivity of gold nanocone array and explore their reflection properties, we carried out a BRDF study on the array. As shown in Figure 8, the solid-colored lines represent BRDF at different angles of incidence from 5 to 75°.
Taking 0.405 $\mu$m light as an example, the BRDF of gold nanocone array can reach as low as 0.002 sr$^{-1}$ at incident light of 5°. The BRDF of gold nanocone array increases with both incidence angle and reflectance angle. At each angle of incidence, the BRDF slowly increases as the angle of reflection. Under different incident angle conditions, the BRDF slowly increases with the increase of the incidence angle. If we integrate the BRDF over the all reflection angle and the specular reflection part over a small angle (about ±15° around the maximum), we deduce that the scattered reflectance is predominant, as shown in Table 1. We can observe from Table 1 that the reflectance of the gold nanocone array is mostly scattered reflectance. Indeed, the light trapping effect is less efficient at high angles. [15, 21, 27–29] Except for the condition of incident light of 5°, the BRDF shows a weak peak near the specular reflection angle, and there is no significant specular reflection overall, indicating good Lambertian properties. This phenomenon indicates that the reflection of the gold nanocone array is primarily caused by non-specular scattering rather than specular reflection(see Table 1). This behavior is in stark contrast to that observed in light-absorbing materials with similar structures, such as black silicon, which demonstrate significant specular reflection[15]. When incident light approaches the perpendicular angle, a distinct peak near $\theta_r = 0°$ is obtained because of the occlusion between the light source and detector. The BRDF of the gold nanocone array still retains the same characteristics at other wavelengths.
At different wavelengths, the BRDF increases with increasing wavelength. As the wavelength increases, the BRDF value under the incidence of 5° also increases, this corresponds to the trend of increasing reflectance with longer wavelengths. At long wavelengths, the BRDF increases with both the incidence angle and reflectance angle. **Furthermore, the difference between BRDF values at the specular reflection angle and nonspecular reflection angle at low incidence angle decreases with increasing wavelength.** That is, the proportion of specular reflection of gold nanocone array decreases with increasing wavelength, which means that gold nanocone array has better Lambert characteristics at longer wavelengths.

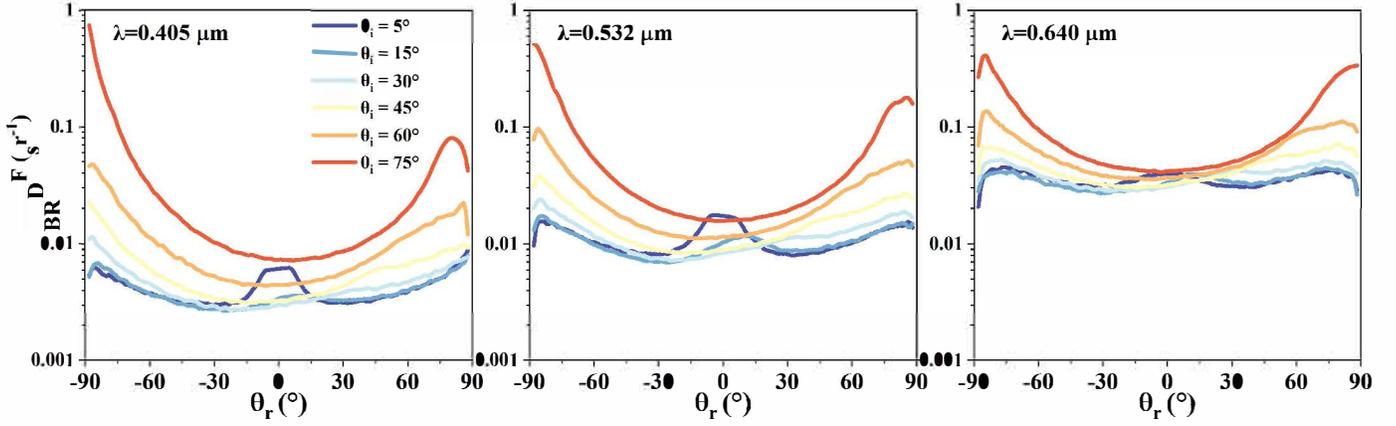

Figure 8: BRDF measurement of gold nanocone array at wavelengths of 405 nm, 532 nm, 640 nm at an angle of incidence of 5 to 75°.

Table 1: Percentage of specular reflectance of the total reflectance.

| $R_{Spec}(\%)$ | 0.405 $\mu m$ | 0.532 $\mu m$ | 0.640 $\mu m$ |
| --- | --- | --- | --- |
| 5° | 21.62 | 22.96 | 18.07 |
| 15° | 14.71 | 17.78 | 18.06 |
| 30° | 14.10 | 16.02 | 17.66 |
| 45° | 15.89 | 17.54 | 19.05 |
| 60° | 18.65 | 20.81 | 23.18 |

## 3 conclusions

In this work, we first determined the optimal geometric parameter of the gold nanocone array using FDTD simulation. Then, we fabricated a wide angle light absorber of the gold nanocone array on the basis of the ion track method. Finally, we experimentally investigated the total absorptivity and BRDF. The gold nanocone array exhibits a high absorption properties and wide angles. FDTD simulations reveal that adjusting the geometric parameters of the gold nanocone array can result in an expanded wavelength range and enhanced absorptivity, and achieve the highest absorptivity of 0.99. Under the guidance of FDTD simulation, we prepared a gold nanocone array with the highest absorptivity measured was reached 0.98, which was basically consistent with the FDTD simulation results. Moreover, the gold nanocone array has good characteristics of light absorber in a wide angle range.
Gold nanocone array have a very low BRDF, which can be as low as 0.405 $mum$ and 0.002 $sr^{-1}$ at incident light of 5°. The BRDF of the gold nanocone array increases as the angles of incidence, angles of reflectance, and wavelength increase. The weak peak near the specular reflection angle indicates that the reflections are primarily due to non-specular reflections. Additionally, the gold nanocone array demonstrates good Lambertian characteristics, particularly at longer wavelengths. This study of absorptivity and BRDF is crucial to comprehending the absorption and space scattering characteristics of gold nanocone array, will contribute to further understanding of the absorption/reflection mechanism of the gold nanocone array and further improving the absorptivity of light absorber.

# 4 Experimental section

*Fabrication of gold nanocone array*: In this work, we employ the ion track method to fabricate gold nanocone array. First, we irradiated a 30 $\mu m$ thick polycarbonate foil with an ion flux of 1e8 ions/cm$^2$. Next, we performed asymmetric etching using a 5 M NaOH solution to etch one side of the ion track membrane prepared in the first step. Subsequently, gold nanocones were deposited in conical pores applying an impulse voltage sequence (+1.1 V for 15 seconds, −0.2 V for 5 seconds) while using an aqueous solution of Na$_3$Au(SO$_3$)$_2$ at a concentration of 75 g/L. Finally, we completely remove the membrane template using dichloromethane (CH$_2$Cl$_2$). With this method, we fabricated a gold nanocone array with a length of approximately $\tilde{2}2$ $\mu m$ and a density of $1\times10^8$ cons/cm$^2$.

*Finite-Difference Time-Domain (FDTD)*: In this study, a three-dimensional model of the gold nanocone array is constructed to simulate the greatest extent by using the nanocone structure generated in FDTD Solutions software. The simulation model is show in Figure 1. The x- and y-directions are set to periodic boundary conditions to simplify the model and improve the computational efficiency. Meanwhile, the z-direction is set to perfectly match the boundary conditions. A plane wave light source is positioned above the nanocone array, and the transmittance detectors are arranged at the top and bottom of the nanocone array in the z-direction to detect the reflected and transmitted light to calculate absorption, the absorptivity (A) was calculated using the formula $A = 1 - R - T$. The material optical parameter of the nanocone is "Gold-Johnson and Christy"[30].

*Characterization*: SEM images were taken with a FEI NanoSEM 450 field emission scanning electron microscopy. A Perkin-Elmer Lambda 1050+ UV-VIS-NIR spectrophotometer equipped with an integrating sphere was used to measure the hemisphere reflectance (R) and transmittance (T), and since T was so low that it was negligible, the absorptivity (A) was actually calculated using the formula $A = 1 - R$. The samples were placed at an 8° angle relative to the incident beam direction. A Super Optics Technology MWS multiple wavelength scatterometer was used to measure the BRDF. The MWS multiple wavelength scatterometer utilizes the relative measurement method and the calibration is transmitted through the NIST-traceable Spectralon standard sheet in the United States. The BRDF of the target is obtained by comparative measurement of the target sample and the standard sample. The BRDF measuring device has a measurement wavelength range of 320-1550 nm, an incident zenith angle range of -90° to 90°, a detection zenith measurement angle range of -90° to 90°, an azimuth angle of 0 to 360°, an angular resolution of 0.001°, a sensitivity better than $10^{-7}$ sr$^{-1}$, an accuracy of better than ±3 %, and a measurements repeatability better than ±2 %.

**Acknowledgements**
The author would like to thank the nuclear membrane team of the Advanced Energy Science and Technology Guangdong Laboratory and the Nanomaterials Group of the Institute of Modern Physics, Chinese Academy of Sciences, for their help in the experimental and artical work, the Lanzhou Heavy Ion Research Facility (HIRFL), and the staff of the accelerator department for providing high-quality beams. This work was supported by the Guangdong Provincial Basic and Applied Basic Research Foundation (Grant No. 2023A1515140068).